\documentclass[sigconf,natbib=true]{acmart}

\usepackage{textcomp}
\usepackage{amsmath}
\usepackage{booktabs}
\usepackage{graphicx}
\usepackage{multirow}
\usepackage{makecell}
\usepackage{enumitem}
\usepackage{placeins}
\usepackage{xspace}
\usepackage{algorithm}
\usepackage{algpseudocode}
\usepackage{color, colortbl}
\usepackage[capitalize,noabbrev]{cleveref}

\newcommand{\brave}{BRAVE\xspace}

\newcommand{\braveSimple}{BRAVE$^{\mathrm{Simple}}$\xspace}

\newcommand{\braveUltra}{BRAVE$^{\mathrm{Ultra}}$\xspace}

\copyrightyear{2027}
\acmYear{2027}
\setcopyright{cc}
\setcctype{by}
\acmConference[WSDM '27]{Proceedings of the Twentieth ACM International Conference on Web Search and Data Mining}{February 15--19, 2027}{Hong Kong, Hong Kong}
\acmBooktitle{Proceedings of the Twentieth ACM International Conference on Web Search and Data Mining (WSDM '27), February 15--19, 2027, Hong Kong, Hong Kong}
\acmDOI{10.1145/3862322.3863445}
\acmISBN{979-8-4007-3117-4/2027/02}

\begin{document}

\title[Budgeted Reflection Allocation for Recommendation Agents]{Reflection Is Not Always Worth It: Budgeted Reflection Allocation for Recommendation Agents}

\author{Jinfeng Xu}
\email{jinfeng.xu@ubc.ca}
\authornote{Corresponding author.}
\affiliation{%
  \institution{The University of British Columbia}
  \city{Vancouver}
  \country{Canada}}

\author{Zheyu Chen}
\email{zyueee.chen@connect.polyu.hk}
\affiliation{%
  \institution{The Hong Kong Polytechnic University}
  \city{Hong Kong}
  \country{Hong Kong}}

% \author{Shuo Yang}
% \email{shuoyang.ee@gmail.com}
% \affiliation{%
%   \institution{The University of Hong Kong}
%   \city{HongKong SAR}
%   \country{China}}

% \author{Jinze Li}
% \email{lijinze-hku@connect.hku.hk}
% \affiliation{%
%   \institution{The University of Hong Kong}
%   \city{HongKong SAR}
%   \country{China}}

\author{Ziyue Peng}
\email{zpengbe@connect.ust.hk}
\affiliation{%
    \institution{The Hong Kong University of Science and Technology
}  
  \city{Hong Kong}
  \country{Hong Kong}}

\author{Jianheng Tang}
\email{tangentheng@stu.pku.edu.cn}
\affiliation{%
    \institution{Peking University}  
    \city{Beijing}   
    \country{China}}

\author{Zheng Lin}
\email{zhenglin@ieee.org}
\affiliation{%
  \institution{University of Luxembourg}
  \city{Luxembourg}
  \country{Luxembourg}}

\author{Wenhao Yuan}
\email{wenhao.yuan@connect.hku.hk}
\affiliation{%
  {\institution{The University of Hong Kong}}
  \city{Hong Kong}
  \country{Hong Kong}}

\author{Jian Chen}
\email{ccccccj03@connect.hku.hk}
\affiliation{%
  {\institution{The University of Hong Kong}}
  \city{Hong Kong}
  \country{Hong Kong}}

\author{Shujie Li}
\email{u3012850@connect.hku.hk}
\affiliation{%
  {\institution{The University of Hong Kong}}
  \city{Hong Kong}
  \country{Hong Kong}}

\author{Victor Leung}
\authornotemark[1]
\email{vleung@ieee.org}
\affiliation{%
  \institution{The University of British Columbia
}
  \city{Vancouver}
  \country{Canada}}

\renewcommand{\shortauthors}{Jinfeng Xu et al.}

\begin{abstract}
Language-model agents are increasingly used as high-level controllers for recommender systems, where they inspect user profiles and candidate lists to decide how multiple recommender tools should be combined. Most recommendation-agent frameworks treat reflection as an always-on operation: every sampled user receives LLM-based recommender evaluation at every agent-training epoch. We argue that reflection is useful but not uniformly worth its cost, raising a central question for agentic recommendation: \emph{which users should receive LLM reflection under a limited budget?} This paper introduces \brave (\textbf{B}udgeted \textbf{R}eflection \textbf{A}llocation via \textbf{V}alue \textbf{E}stimation), a framework that treats recommendation-agent training as selective reflection under a budget. \brave estimates user-level reflection value from cheap recommender-side uncertainty before calling the LLM, then allocates training-time reflection to estimated high-value users while preserving full-catalog ranking and full inference reranking. The final allocator is gated: ranking fragility determines whether reflection has room to change the top list, and cross-tool disagreement is used only as a secondary signal under this gate. A rank-flip analysis explains why low-margin users are more likely to benefit from memory updates and why ungated disagreement can measure conflict rather than value. Experiments on full-ranking recommendation benchmarks show that \brave matches or improves a uniform-reflection agent while using fewer repeated training-reflection calls, and a controlled calibration study shows that its score tracks realized user-level reflection gain. These results suggest that scalable recommendation agents need policies for allocating reflection, not only stronger reflection modules.
\end{abstract}




\maketitle

\section{Introduction}
Large language models (LLMs) have become an appealing interface for recommender systems because they can reason over user histories, item text, and intermediate ranking evidence~\cite{bao2023tallrec,geng2022p5,hou2024llmrank,zhang2024generativeagents,zhang2024agentcf,wang2025simulation,xu2026beyond}. Yet direct LLM recommendation is still poorly matched to full-ranking scenarios: LLMs may hallucinate non-existent item IDs, struggle with large catalogs, and incur substantial latency when asked to reason over many candidates~\cite{ji2023hallucination,jiang2025beyond}. A recent line of work therefore casts the LLM as an agent that orchestrates conventional recommenders as external modules~\cite{zhao2024toolrec,yang2026prta,yang2026agentdr}, echoing the broader shift from passive generation to language agents that call external modules and APIs~\cite{yao2023react,schick2023toolformer,wu2024avatar}. In this setting, recommender tools provide valid full-catalog rankings, and the LLM reasons over their exposed evidence.

This architecture raises a question that classical recommendation rarely needs to ask: \emph{how should reflection be allocated?} Existing recommendation agents mainly study prompts, memories, and recommender-tool interfaces, and typically apply reflection to every sampled user once reflection is deemed useful. The policy is simple, but the value of reflection is not uniform. Recent LLM routing and cascade studies make a similar point at the system level: expensive LLM computation should be invoked selectively when cost and quality must be balanced~\cite{chen2023frugalgpt,yue2024llmcascade}. In recommendation agents, some users have sharp and mutually consistent recommender-tool rankings, so the mixture over recommenders is nearly determined before LLM feedback. Others have flat score margins or conflicting recommender evidence, where a small change in personalized tool-mixture memory can move relevant items near the top of the list. Treating both groups identically conflates \emph{reflection capability} with \emph{reflection allocation}. The distinction matters because LLM judgments are useful but noisy: unnecessary updates can consume calls and perturb reliable mixture memories, especially when LLM evaluators exhibit order or self-evaluation biases~\cite{wang2024llmnotfair,xu2024selfbias}.

We study reflection allocation as a first-class problem for recommender agents. The goal is to separate two roles that are often entangled. The \emph{reflector} defines how LLM feedback changes a user's tool-mixture memory, following the broader literature on language-agent reflection and self-feedback~\cite{shinn2023reflexion,madaan2023selfrefine,zhou2024lats}; the \emph{allocator} decides which users should receive that feedback. Personalized recommender-tool learning agents mainly improve the reflector. This paper asks whether the allocator can be built from recommender evidence before any LLM call is made. The objective is not only to reduce cost, but also to improve agent training under a limited reflection budget.

We instantiate this idea as \brave, short for \textbf{B}udgeted \textbf{R}eflection \textbf{A}llocation via \textbf{V}alue \textbf{E}stimation. Before invoking the LLM, \brave estimates which users are likely to benefit from reflection and allocates the limited budget to those users. The estimator is gated: margin-based ranking fragility is the primary signal, while cross-tool disagreement is admitted only when the current ranking is sufficiently fragile. This requires no additional model training and no extra LLM calls. \brave is not a new recommender backbone or reflection prompt; it decides when the existing reflection operation should be spent. The recommender tools still rank the whole catalog, and inference-time reranking remains unchanged, isolating the paper's central claim that reflection should be allocated rather than automatically spent.

The design also clarifies a common intuition about uncertainty. Cross-tool disagreement sounds like a natural reason to reflect: if recommender tools disagree, the LLM should decide which recommender to trust. This intuition is related to classic query-by-committee active learning, where committee disagreement identifies informative examples~\cite{seung1992query}. In recommendation agents, however, disagreement is not always value. It may indicate a recoverable conflict where one recommender exposes the relevant neighborhood, or it may indicate that all recommenders are locally unreliable. In the latter case, reflection spends calls reallocating weight among weak alternatives. We therefore distinguish \emph{ranking fragility}, which measures whether a memory update can change the decision, from \emph{cross-tool disagreement}, which measures whether recommender tools produce different evidence. \brave operationalizes this distinction with a gate: margin uncertainty first identifies users near a ranking decision boundary~\cite{tong2001svm}, and disagreement is then used only as a conditional secondary signal.

We evaluate \brave on full-ranking recommendation benchmarks. Across datasets and reflection budgets, \brave preserves the accuracy benefits of uniform reflection while using fewer LLM calls, and its strongest budget setting improves top-ranked quality over a full-reflection agent baseline. A controlled branch comparison further shows that the allocation score is correlated with realized user-level reflection gain, rather than only improving aggregate metrics after the fact. We also find that disagreement is useful only when conditioned on ranking fragility. The broader message is that recommendation agents should optimize not only \emph{how} they reflect, but also \emph{when} reflection is warranted and which uncertainty signals should control that decision.

% Our contributions are:
% \begin{itemize}[leftmargin=*]
%     \item We introduce reflection allocation as a budget-constrained learning problem for full-ranking recommendation agents.
%     \item We propose \brave, a gated reflection-allocation framework that estimates user-level reflection value from recommender-side self-doubt before any training-time LLM call is made.
%     \item We provide a rank-flip and budget-stability analysis explaining why margin uncertainty should gate reflection opportunity and why disagreement should be used conditionally.
%     \item We provide controlled full-ranking experiments, realized-gain calibration, and allocator diagnostics showing that value-aware reflection allocation improves the cost-quality trade-off over uniform reflection.
% \end{itemize}

\section{Related Work}
\label{sec:related}

\subsection{LLMs for Recommendation}
LLM-based recommendation has moved from semantic scoring toward reasoning over user histories, item text, and ranking evidence. Early approaches formulate recommendation as point-wise preference prediction, prompt-based scoring, or zero-shot reranking~\cite{geng2022p5,bao2023tallrec,hou2024llmrank,xu2026iimrec,xu2026ntcf}. They exploit textual knowledge but often operate on restricted candidate sets and remain weakly coupled with full-catalog collaborative signals. A second line aligns collaborative representations with language-model spaces, enabling LLMs to consume user/item embeddings or generate recommendation outputs~\cite{zhang2024agentcf,jiang2025beyond}. More recent work further emphasizes preference reasoning beyond shallow semantic matching; for example, ThinkRec uses reasoning activation and personalized expert fusion to improve LLM-based recommendation~\cite{yu2026thinkrec}.

Together, these studies show that LLMs can enrich recommendation with semantic and reasoning ability, but they also expose a deployment tension: direct LLM recommenders must generate valid item identifiers, rerank limited candidates, or pay substantial inference and training costs. \brave takes a complementary route. It keeps full-catalog scoring inside conventional recommenders and uses the LLM only for agentic reflection and final reranking. The key question therefore shifts from making the LLM recommend directly to deciding when its costly reflection is worth invoking during agent training.

\subsection{Language Agents with External Modules}
Language agents often extend LLMs beyond pure text generation by connecting them to external knowledge sources, symbolic modules, APIs, or specialist models~\cite{karpas2022mrkl,shen2023hugginggpt}. This line is commonly described as LLM tool use, but the ``tools'' in that literature range from search engines and calculators to code interpreters and domain models. Representative systems interleave reasoning and action~\cite{yao2023react}, learn module invocation from self-supervised demonstrations~\cite{schick2023toolformer}, improve API calling with retrieval-aware training~\cite{patil2024gorilla}, and optimize module-use trajectories through contrastive or trajectory-level reasoning~\cite{wu2024avatar}. These works establish external-module access as a way to mitigate missing knowledge, limited exact computation, and unreliable direct generation.

Recommendation agents instantiate this paradigm where output validity and catalog scale are central. Rather than asking an LLM to generate item identifiers, module-based recommendation agents use conventional recommenders to produce valid full-catalog rankings and let the LLM reason over recommender evidence. ToolRec studies LLM-empowered selection among recommender modules~\cite{zhao2024toolrec}, AgentDR uses LLM agents for dynamic recommendation~\cite{yang2026agentdr}, and PRTA learns personalized tool-mixture memories through local recommender evaluation, recommender comparison, and ranking comparison~\cite{yang2026prta}. \brave studies a complementary dimension: given the same reflection operation, which users should receive LLM feedback under a budget? This separates the reflector from the allocator and treats user-level reflection as a selective training decision.

\subsection{Cost-Aware Reasoning and Uncertainty-Based Selection}
\brave is also related to cost-aware reasoning and selective computation. LLM routing and cascade methods invoke expensive reasoning only when cheaper models or intermediate states are insufficient~\cite{chen2023frugalgpt,yue2024llmcascade}. Adaptive computation, early exiting, and selective prediction similarly allocate computation or coverage according to confidence~\cite{graves2016act,xin2020deebert,geifman2017selective,geifman2019selectivenet}. Their shared principle is that expensive computation should follow expected value rather than a uniform schedule. Reflection allocation differs from model routing or abstention because the decision is not which model answers a query, but whether an LLM should update a user's personalized tool-mixture memory during training. Its value is measured by downstream top-$K$ ranking after the memory update, making recommender-side evidence especially useful before any LLM call is made. This connects \brave to active learning: uncertainty sampling and margin selection prioritize examples near decision boundaries~\cite{lewis1994uncertainty,tong2001svm,settles2009activelearning}, while query-by-committee uses disagreement as an informativeness signal~\cite{seung1992query}. In recommendation agents, however, disagreement can indicate either complementary evidence or uniformly weak recommenders, so \brave gates disagreement by ranking fragility instead of treating it as an unconditional objective.

\section{Problem Setup}
\label{sec:prelim}

We consider an agentic full-ranking setting in which conventional recommenders produce valid catalog-wide scores and an LLM agent maintains personalized tool-mixture memories. The key resource is training-time reflection: it can improve the mixture memory, but each call is expensive. We therefore formulate reflection as a budgeted allocation problem that \brave must approximate before making LLM calls. Throughout this paper, unless we explicitly discuss external-module language-agent literature, a \emph{tool} refers to a pretrained full-catalog recommender. A \emph{tool pool} is a set of such recommenders, and a \emph{tool-mixture memory} denotes the user-specific weights used to combine their scores.

\subsection{Agentic Full-Ranking Recommendation}
Let $\mathcal{U}$ and $\mathcal{I}$ denote users and items. For each user $u$, a recommender must rank all items in $\mathcal{I}$ and is evaluated on held-out future interactions. We assume a set of pretrained recommender tools $\mathcal{T}=\{1,\ldots,M\}$. Each recommender tool $m$ returns a full score vector $\mathbf{s}_{u}^{m}\in\mathbb{R}^{|\mathcal{I}|}$ over the complete item catalog. The formulation is agnostic to the concrete recommender family: the pool may include graph-based, sequential, collaborative-filtering, diffusion-based, or other full-catalog recommenders, as long as they expose scores or top-ranked lists.

The agent maintains a personalized tool-mixture memory $\mathbf{w}_{u}\in\mathbb{R}^{M}$ for each user and aggregates recommender-tool scores as:
\begin{equation}
\mathbf{s}_{u}=\sum_{m=1}^{M} w_{u,m}\mathbf{s}_{u}^{m}.
\label{eq:aggregate}
\end{equation}

The LLM therefore does not generate item IDs or rank the full catalog directly. Its role is to update the tool-mixture memory by judging textual evidence from user profiles, recent interactions, and top-ranked candidate lists.

\subsection{Reflection as Costly Agentic Computation}
Such an agent can update memory with three complementary signals: LLM-based local recommender evaluation, LLM-based global recommender comparison, and a non-LLM ranking comparison~\cite{yang2026prta}. Local evaluation judges each recommender tool independently, while global comparison chooses the best recommender among candidate lists. With $M$ recommender tools, $n$ users, and $t$ epochs, a uniform full-reflection policy uses $n$ profile calls, $(M+1)nt$ training reflection calls, and $n$ inference reranking calls, for $(M+1)nt+2n$ total LLM calls. We define a reflection policy $\pi(u)\in\{0,1\}$ that decides whether user $u$ receives training-time LLM reflection. Given a budget $B\in(0,1]$, the policy must satisfy
\begin{equation}
\sum_{u\in\mathcal{U}_{0}}\pi(u)\leq B|\mathcal{U}_{0}|,
\label{eq:budget_constraint}
\end{equation}
where $\mathcal{U}_{0}\subseteq\mathcal{U}$ is the sampled-agent window used by the LLM-agent stage. Following PRTA~\cite{yang2026prta}, this window controls the number of user-level LLM calls; it does not restrict recommender training or the item catalog. The learning objective is to maximize ranking utility after the memory update under this budget:
\begin{equation}
\max_{\pi}\;\mathcal{R}(\mathbf{W}^{+}_{\pi})-\lambda C(\pi),
\label{eq:objective}
\end{equation}
where $\mathcal{R}$ is a top-$K$ ranking metric, $\mathbf{W}^{+}_{\pi}$ denotes the post-update tool-mixture memories induced by policy $\pi$, $C(\pi)$ is the realized LLM-call cost, and $\lambda$ controls the cost-accuracy trade-off. Directly optimizing \cref{eq:objective} is unacceptable before making LLM calls, so the central challenge is to build a cheap pre-reflection surrogate for the value of reflection.

\section{Budgeted Reflection Allocation}
\label{sec:method}
\begin{figure*}
    \centering
    \includegraphics[width=0.95\linewidth]{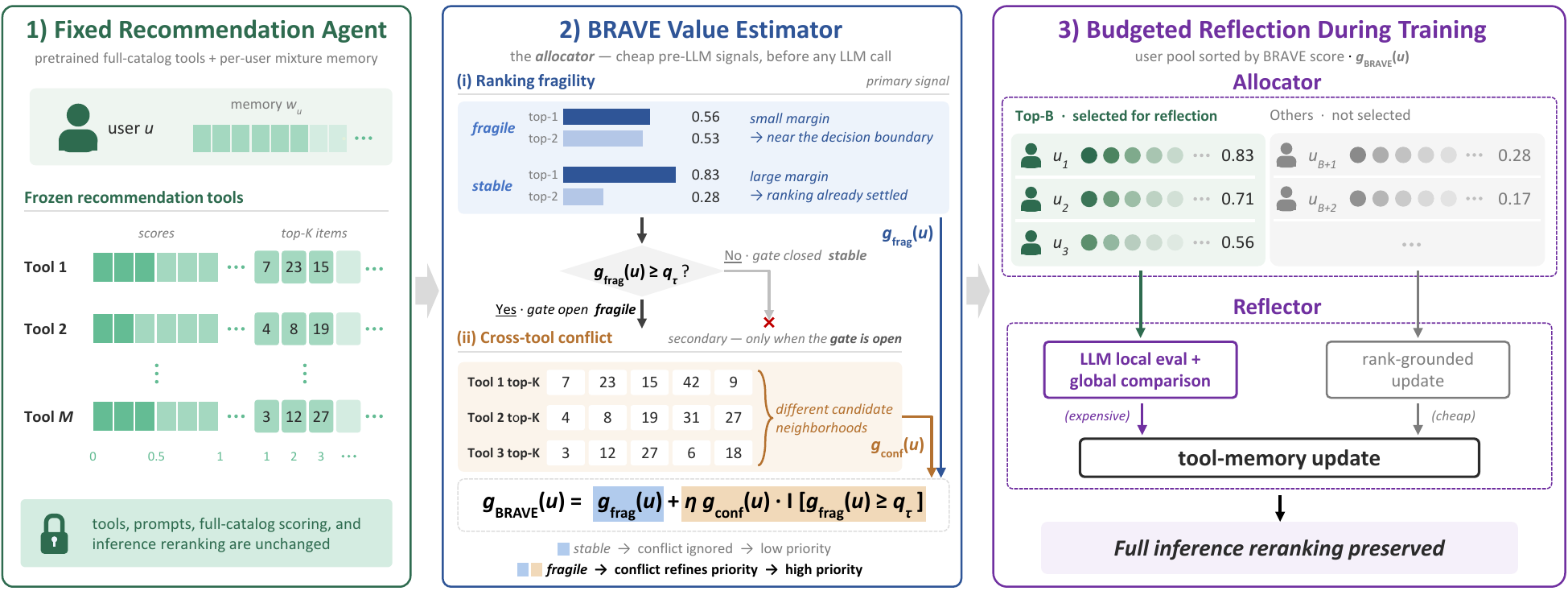}
    \vskip -0.15in
    \caption{The overall structure of BRAVE.}
    \label{fig:framework}
    \vskip -0.1in
\end{figure*}

This section presents BRAVE\footnote{Code can be found at: \href{https://github.com/Jinfeng-Xu/BRAVE}{https://github.com/Jinfeng-Xu/BRAVE}.} as a budgeted reflection allocator on top of a fixed recommendation agent. \brave keeps the recommender tools, reflection prompts, and inference reranking unchanged, and estimates before training which users are worth LLM reflection. The method consists of a reflection-value surrogate, a gated uncertainty estimator, and a budgeted memory-update policy.

\subsection{Reflection Value Surrogate}
\brave treats the base agent as a fixed reflector plus a budgeted allocator. The \textit{reflector} defines the memory update once LLM feedback is available; the \textit{allocator} decides whether that feedback is worth its cost. For a user $u$, let $\mathbf{W}^{+}_{u,\mathrm{LLM}}$ denote post-update mixture memories with LLM local evaluation and global comparison, and let $\mathbf{W}^{+}_{u,\mathrm{rank}}$ denote mixture memories with only the non-LLM rank-grounded update. We define the latent value of reflection as:
\begin{equation}
V(u)=\mathbb{E}\!\left[\mathcal{R}(\mathbf{W}^{+}_{u,\mathrm{LLM}})-\mathcal{R}(\mathbf{W}^{+}_{u,\mathrm{rank}})\mid \mathcal{E}_{u}\right],
\label{eq:value}
\end{equation}
where $\mathcal{R}(\cdot)$ is a top-$K$ ranking utility such as Recall or NDCG, and $\mathcal{E}_{u}$ denotes pre-reflection evidence, including normalized recommender-tool scores, top-ranked lists, and the current tool-mixture memory. Since $V(u)$ depends on unobserved LLM feedback and cannot be computed before the call, \brave estimates it with a recommender-side score computed from saved recommender outputs.

The core design choice is that not all uncertainty should control reflection equally. \brave uses two signals with different roles. \emph{Ranking fragility} measures whether a bounded tool-mixture memory update can change top-ranked items. \emph{Cross-tool disagreement} measures whether different recommender tools expose different candidate neighborhoods. The former is the primary signal; the latter is used only when the current ranking is fragile enough for reflection to plausibly matter.

\subsection{Ranking Fragility}
For each recommender tool $m$, let $\tilde{\mathbf{s}}_{u}^{m}$ be the score vector after per-user min-max normalization over the inspected candidate union, so that scores from different recommenders are comparable. Let $\tilde{s}_{u,(r)}^{m}$ denote the $r$-th largest entry in $\tilde{\mathbf{s}}_{u}^{m}$. We compute a top-margin uncertainty:$
\Delta_{u}^{m}=\tilde{s}_{u,(1)}^{m}-\tilde{s}_{u,(2)}^{m}$,
where a small $\Delta_{u}^{m}$ means that recommender tool $m$ is not confident about its preferred item. A memory update can therefore more easily change the final top-ranked set. We aggregate this signal as:
\begin{equation}
g_{\mathrm{frag}}(u)=-\frac{1}{M}\sum_{m=1}^{M}\Delta_{u}^{m}.
\label{eq:uncertainty}
\end{equation}

A larger $g_{\mathrm{frag}}(u)$ indicates that the current ranking is more fragile, since the top-ranked items have smaller average score gaps within the recommender pool. This signal serves as a low-cost proxy for aggregate rank-boundary sensitivity: it can be computed before reflection from cached recommender scores, without simulating LLM feedback or enumerating possible mixture-memory updates. With cached top-$K$ outputs, its cost is $O(M|\mathcal{U}_{0}|K)$; sorting is required only when candidate lists are not already cached. \cref{sec:rankflip} motivates why low-margin users offer greater rank-flip opportunity, while \cref{sec:value-calibration,sec:ablation} compare this proxy with aggregate-margin alternatives.

\subsection{Cross-Tool Disagreement}
Ranking fragility alone may miss cases where each recommender tool is individually confident but recommenders expose different item neighborhoods. \brave computes a disagreement score over the top-$K$ set $\mathcal{L}_{u}^{m}\subseteq\mathcal{I}$ returned by each recommender tool $m$:
\begin{equation}
g_{\mathrm{conf}}(u)=1-\frac{2}{M(M-1)}\sum_{a<b}
\frac{|\mathcal{L}_{u}^{a}\cap\mathcal{L}_{u}^{b}|}{|\mathcal{L}_{u}^{a}\cup\mathcal{L}_{u}^{b}|}.
\label{eq:disagreement}
\end{equation}

High cross-tool disagreement indicates that personalized recommender weighting may matter because different recommenders expose different candidate regions. However, disagreement is not the same as value: it can arise because one recommender captures the relevant neighborhood, but it can also arise because all recommenders are unreliable for the current user. This motivates using disagreement as a conditional signal rather than as an unconditional additive objective.

\subsection{Gated Value Estimator}
\brave combines the two signals with a gated estimator:
\begin{equation}
g_{\mathrm{BRAVE}}(u)=g_{\mathrm{frag}}(u)+
\eta\,g_{\mathrm{conf}}(u)\cdot \mathbb{I}[g_{\mathrm{frag}}(u)\geq q_{\tau}],
\label{eq:gated}
\end{equation}
where $q_{\tau}$ is the $\tau$-quantile of $\{g_{\mathrm{frag}}(u):u\in\mathcal{U}_{0}\}$ over the sampled-agent window, $\eta\geq0$ controls the strength of the conditional disagreement term, and $\mathbb{I}[\cdot]$ is the indicator function. The gate first identifies users whose rankings are near a decision boundary, then uses cross-tool disagreement to refine their priority. Disagreement can be useful when reflection can plausibly change the learned recommender mixture, but noisy when the ranking is already stable. In the experiments, margin-only and ungated additive scores are allocator ablations of this estimator.

\subsection{Budgeted Reflection Policy}
Given budget $B$, \brave selects the users with the largest estimated reflection value:
\begin{equation}
\mathcal{U}_{B}=\operatorname{Top}_{\lceil B|\mathcal{U}_{0}|\rceil}\{g_{\mathrm{BRAVE}}(u):u\in\mathcal{U}_{0}\}.
\label{eq:budget}
\end{equation}

During agent training, LLM local-evaluation and global-comparison updates are applied only to $u\in\mathcal{U}_{B}$. Users outside $\mathcal{U}_{B}$ still receive the rank-grounded comparison update, so every user can improve its tool-mixture memory without incurring LLM reflection. At epoch $e$, the update rule can be written as:
\begin{equation}
\mathbf{w}_{u}^{e+1}=\mathbf{w}_{u}^{e}
+\gamma_{\mathrm{rank}}\mathbf{r}_{u}^{e}
+\pi_{B}(u)(\alpha\mathbf{l}_{u}^{e}+\beta\mathbf{c}_{u}^{e}),
\label{eq:update}
\end{equation}
where $\mathbf{w}_{u}^{e}\in\mathbb{R}^{M}$ is the tool-mixture memory before epoch $e$, $\mathbf{r}_{u}^{e}\in\mathbb{R}^{M}$ is the non-LLM rank-comparison signal, and $\mathbf{l}_{u}^{e},\mathbf{c}_{u}^{e}\in\mathbb{R}^{M}$ are the LLM local-evaluation and global-comparison signals. The coefficients $\gamma_{\mathrm{rank}}$, $\alpha$, and $\beta$ are learning rates for the three update sources. The budget indicator $\pi_{B}(u)=\mathbb{I}[u\in\mathcal{U}_{B}]$ is fixed after allocation, so LLM reflection affects only selected users while the rank-grounded update remains available to all users.

\begin{algorithm}[t]
\caption{\brave training-time reflection allocation}
\label{alg:brave}
\begin{algorithmic}[1]
\Require User window $\mathcal{U}_{0}$, recommender tools $\mathcal{T}$, budget $B$, epochs $t$, disagreement weight $\eta$, gate quantile $\tau$
\State Obtain normalized recommender-tool scores $\{\tilde{\mathbf{s}}_{u}^{m}\}_{u,m}$ and top lists $\{\mathcal{L}_{u}^{m}\}_{u,m}$
\For{each user $u\in\mathcal{U}_{0}$}
    \State Compute ranking fragility $g_{\mathrm{frag}}(u)$ from top-score margins by \cref{eq:uncertainty}
    \State Compute cross-tool disagreement $g_{\mathrm{conf}}(u)$ from top-list overlap by \cref{eq:disagreement}
\EndFor
\State Compute the gate threshold $q_{\tau}$ from $\{g_{\mathrm{frag}}(u):u\in\mathcal{U}_{0}\}$
\For{each user $u\in\mathcal{U}_{0}$}
    \State Compute $g_{\mathrm{BRAVE}}(u)$ by \cref{eq:gated}
\EndFor
\State Select $\mathcal{U}_{B}=\operatorname{Top}_{\lceil B|\mathcal{U}_{0}|\rceil}\{g_{\mathrm{BRAVE}}(u)\}$
\For{$e=1,\ldots,t$}
    \For{each minibatch of users}
        \State Apply non-LLM rank-grounded update to all users
        \State Query LLM local evaluation and global comparison only for users in $\mathcal{U}_{B}$
        \State Update tool-mixture memories with \cref{eq:update}
    \EndFor
\EndFor
\State Keep the original full-catalog scoring and inference-time reranking protocol unchanged
\end{algorithmic}
\end{algorithm}

\subsection{Cost and Design Properties}
Let $n=|\mathcal{U}_{0}|$ be the number of sampled agents, $t$ be the number of agent-training epochs, and $M$ be the number of recommender tools. Under a full-reflection agent setting, each user requires one profile-generation call, each reflected user requires $M+1$ training-time reflection calls per epoch ($M$ local recommender evaluations and one global comparison), and each user receives one inference-time reranking call. \brave therefore uses:
\begin{equation}
C(B,t,M)=n+(M+1)Bnt+n=(M+1)Bnt+2n.
\label{eq:cost}
\end{equation}

LLM calls when profile generation and inference reranking are kept for all users. Compared with the uniform full-reflection policy $C_{\mathrm{full}}(t,M)=(M+1)nt+2n$, the relative call ratio is:
\begin{equation}
\frac{C(B,t,M)}{C_{\mathrm{full}}(t,M)}
=\frac{(M+1)Bt+2}{(M+1)t+2}.
\label{eq:cost_ratio}
\end{equation}

For the three-tool Ultra pool, this reduces to $(4Bt+2)/(4t+2)$; for a four-recommender pool, the reflected training term becomes $5Bnt$. Thus \brave reduces only repeated training reflection while preserving inference capability. It is \emph{tool-agnostic}, \emph{prompt-preserving}, and \emph{fail-soft}: when $B=1$, it recovers uniform full reflection.

\section{Theoretical Analysis}
\label{sec:analysis}

This section analyzes why allocation can help even when the reflector is unchanged: when a memory update can affect the final ranking, and why uncertainty signals differ as value proxies.

\subsection{Rank-Flip Opportunity}
\label{sec:rankflip}
Let $\mathbf{z}_{u}(i)=[\tilde{s}_{u}^{1}(i),\ldots,\tilde{s}_{u}^{M}(i)]$ stack the normalized scores of item $i$ across recommender tools, and let:
\begin{equation}
S_u(i)=\mathbf{w}_{u}^{\top}\mathbf{z}_{u}(i),
\end{equation}
be the aggregated score before reflection. Suppose a reflection step changes the tool-mixture memory by $\delta\mathbf{w}_{u}$, yielding $S_u^{+}(i)=(\mathbf{w}_{u}+\delta\mathbf{w}_{u})^{\top}\mathbf{z}_{u}(i)$. If normalized scores lie in $[0,1]$ and $\|\delta\mathbf{w}_{u}\|_1\leq\rho$, then for any two items $i$ and $j$,
\begin{equation}
| (S_u^{+}(i)-S_u^{+}(j))-(S_u(i)-S_u(j)) |\leq 2\rho.
\label{eq:perturb_bound}
\end{equation}

Therefore, if the current top item $i_{(1)}$ satisfies
\begin{equation}
S_u(i_{(1)})-S_u(j)>2\rho,\quad \forall j\neq i_{(1)},
\label{eq:rankflip}
\end{equation}
then no reflection update with $\ell_1$ size at most $\rho$ can change the top-ranked item.
The same perturbation argument applies to the top-$K$ boundary: if every item inside the current top-$K$ list is separated from every item outside the list by more than $2\rho$, then a bounded memory update cannot change the top-$K$ set. Thus the relevant opportunity for Recall@K and NDCG@K is not uncertainty in isolation, but uncertainty near a ranking boundary.

\paragraph{Implication.}
\cref{eq:rankflip} shows that large-margin users are insensitive to bounded memory updates: even good reflection has little room to change their ranking. Low-margin users are closer to a decision boundary, so modest memory correction can change which recommender neighborhood dominates. The argument concerns aggregate score margin under $\mathbf{w}_{u}$; \brave uses a cheaper per-recommender margin as a pre-reflection proxy. This proxy is exact only when mixture memories are roughly balanced and recommender-level boundaries align with aggregate boundary, so we compare it with aggregate-margin and top-$K$ boundary baselines in \cref{sec:ablation} and validate it against realized gain in \cref{sec:value-calibration}.

\subsection{Budgeted Selection as Value Approximation}
\label{sec:value-approximation}
Let $V(u)$ denote the unobserved reflection value in \cref{eq:value}, and suppose the allocator observes a noisy surrogate
\begin{equation}
g(u)=V(u)+\epsilon_u,
\end{equation}
where $\epsilon_u$ is surrogate error. Under a budget $B$, the optimal allocation selects the top $k=\lceil B|\mathcal{U}_{0}|\rceil$ users by $V(u)$, while \brave selects the top $k$ users by $g(u)$. If $\Delta_k$ is the value gap between the $k$-th and $(k+1)$-th users under the true values, then every user whose surrogate error satisfies $|\epsilon_u|<\Delta_k/2$ preserves its side of the budget boundary. Thus the allocation is stable when the value signal has a clear separation near the budget cutoff.

\paragraph{Implication.}
The surrogate need not rank all users perfectly; it must avoid replacing high-value boundary users with low-value ones near the budget cutoff. An added uncertainty signal can therefore hurt if it changes cutoff users without improving true reflection value. We examine this boundary-swap effect in \cref{sec:allocator-analysis}.

\subsection{Why Disagreement Requires Calibration}
\label{sec:disagreement-calibration-theory}
Cross-tool disagreement is an intuitive uncertainty signal, but it is not identical to reflection value. High disagreement means that recommender tools retrieve different candidate neighborhoods:
\begin{equation}
g_{\mathrm{conf}}(u)\uparrow \quad \Longrightarrow \quad
\{\mathcal{L}_{u}^{m}\}_{m=1}^{M}\;\text{have low overlap}.
\end{equation}

This can be beneficial when one recommender exposes the relevant neighborhood and reflection can learn to upweight it. It can be harmful when all recommenders are unreliable, because reflection then reallocates weight among weak alternatives. In the latter case, disagreement measures conflict rather than opportunity.

\paragraph{Calibration principle.}
Disagreement should therefore be conditional: ranking fragility measures whether memory updates can change the ranking, while cross-tool disagreement measures whether recommender choice may matter. This motivates the gated estimator in \cref{eq:gated}, whose behavior is tested in \cref{sec:allocator-analysis}.

\section{Experiments}
\label{sec:experiments}

\subsection{Experimental Setup}

\begin{table}[t]
\centering
\caption{Statistics of all datasets.}
\label{tab:dataset-stats}
\small
% % \scriptsize
% \setlength{\tabcolsep}{2.6pt}
% \resizebox{\columnwidth}{!}{
\vskip -0.1in
\begin{tabular}{lccccc}
\toprule
Dataset & Domain & Users & Items & Interactions & Window size  \\
\midrule
Amazon & Product & 1,054 & 1,839 & 35,492 & 160 \\
Yelp & Business & 4,303 & 3,429 & 258,486 & 160 \\
Goodreads & Book & 3,204 & 2,167 & 246,034 & 160 \\
\bottomrule
\end{tabular}
 \vskip -0.1in
\end{table}

\paragraph{Datasets.}
We evaluate on three full-ranking recommendation benchmarks: Amazon~\cite{ni2019justifying}, Yelp~\cite{yelpdataset}, and Goodreads~\cite{wan2018item}. They cover product, local-business, and book recommendation scenarios, with different sparsity levels and textual-description regimes. For each dataset, interactions are ordered by timestamp to construct historical user sequences. Following prior sequential recommendation protocols~\cite{yang2026prta,kang2018sasrec}, we use leave-one-out evaluation. \cref{tab:dataset-stats} summarizes the dataset statistics and agent-evaluation windows.

\paragraph{Baselines.}
We compare \brave against four baseline families. The first comprises conventional full-catalog recommenders: LightGCN~\cite{he2020lightgcn}, SASRec~\cite{kang2018sasrec}, SimpleX~\cite{mao2021simplex}, ENMF~\cite{chen2020enmf}, DiffRec~\cite{wang2023diffrec}, LRURec~\cite{yue2024linear}, FEARec~\cite{du2023fearec}, WGCL~\cite{chen2025squeeze}, and NLGCL~\cite{xu2025nlgcl}. The second contains language-based rankers, BM25~\cite{robertson2009bm25} and LLMRank~\cite{hou2024llmrank}. The third contains RAG-style rerankers, which retrieve candidates from each recommender tool and then use an LLM for reranking. The fourth contains agentic recommender-tool learning baseline PRTA~\cite{yang2026prta}. PRTA~\cite{yang2026prta} performs full LLM reflection for every user, whereas \brave applies the same reflection operation only to budget-selected users. For both PRTA and \brave, $^{\mathrm{Simple}}$ denotes the PRTA tool set, which combines a graph collaborative filter, a sequential recommender, and a matching-based collaborative model: LightGCN~\cite{he2020lightgcn} + SASRec~\cite{kang2018sasrec} + SimpleX~\cite{mao2021simplex}. $^{\mathrm{Ultra}}$ denotes the strongest validation-selected tool combination, whose selection process is analyzed in \cref{sec:tool-pool}. We use GPT-5.6 Luna~\cite{openai2026gpt56} as the default LLM controller and study both closed-source controllers (GPT-4o, GPT-5.6 Luna, GPT-5.6 Terra) and open-source controllers (Phi-4-mini, Qwen3-8B, GPT-OSS-20B) in \cref{sec:controller-robustness}.

\paragraph{Metrics.}
We follow standard top-$K$ evaluation practice in recommendation~\citep{xu2025survey,chen2025hypercomplex}. The best checkpoint is selected by validation Recall@20, and test performance is reported with Recall@10, Recall@20, NDCG@10, and NDCG@20. Reported table values are means over five matched sampled-agent windows.

\paragraph{Implementation details.}
For fairness, \brave follows the PRTA recommender-training and LLM-agent protocol. All recommender tools are trained with a fixed seed on the full training set and rank the entire catalog. The LLM-agent stage is evaluated on five matched 160-user windows per dataset; these windows are shared by PRTA, \brave, and allocator variants and affect only per-user memory optimization/evaluation. Simple uses LightGCN~\cite{he2020lightgcn} + SASRec~\cite{kang2018sasrec} + SimpleX~\cite{mao2021simplex}; Ultra uses the validation-selected NLGCL~\cite{xu2025nlgcl} + WGCL~\cite{chen2025squeeze} + FEARec~\cite{du2023fearec} pool, analyzed in \cref{sec:tool-pool}. We set $\alpha=\beta=0.1$, decay both by $0.5$ after each epoch, and use $k_{\mathrm{eval}}=k_{\mathrm{comp}}=k'_{\mathrm{rank}}=20$.

GPT-5.6 Luna is the default LLM controller, with temperature $0$. Prompts, parsing rules, profile generation, learning rates, and inference reranking are identical between PRTA and \brave; the main comparison changes only training-time reflection allocation. Given budget $B$, \brave selects $\lceil Bn\rceil$ users before training and keeps the set fixed; Random follows the same fixed-set protocol unless explicitly labeled dynamic. Users outside the set still receive rank-grounded updates. Because metrics are averaged over five 160-user windows, Recall values are reported on the corresponding $1/800$ grid. Tool-pool and checkpoint selection use validation interactions only, test items are held out until final evaluation, and $q_{\tau}$ is computed from pre-reflection recommender-tool scores without labels. Unless otherwise specified, $B=0.6$, $\eta=0.25$, and $\tau=0.5$.

\begin{table*}[!t]
\centering
\caption{Recommendation performance comparison of different models. The best and the second-best performances are marked with \textbf{bold} and \underline{underline}, respectively. Results are averaged over five matched 160-user agent-evaluation windows. $^*$ indicates paired significance under the sampled-window evaluation with $p$-value $<$ 0.05.}
\vskip -0.1in
\label{tab:main-results}
\scriptsize
\setlength{\tabcolsep}{2pt}
\resizebox{\textwidth}{!}{
\begin{tabular}{l|cccc|cccc|cccc}
\toprule
\multirow{2}{*}{Method} & \multicolumn{4}{c|}{Amazon} & \multicolumn{4}{c|}{Yelp} & \multicolumn{4}{c}{Goodreads} \\
& R@10 & N@10 & R@20 & N@20 & R@10 & N@10 & R@20 & N@20 & R@10 & N@10 & R@20 & N@20 \\
\midrule
LightGCN~\cite{he2020lightgcn} & 0.0250 & 0.0111 & 0.0375 & 0.0143 & 0.0438 & 0.0216 & 0.0750 & 0.0293 & 0.0625 & 0.0244 & 0.0875 & 0.0309 \\
SASRec~\cite{kang2018sasrec} & 0.0438 & 0.0206 & 0.0813 & 0.0301 & 0.0688 & 0.0413 & 0.1063 & 0.0508 & 0.1000 & 0.0539 & 0.1688 & 0.0714 \\
SimpleX~\cite{mao2021simplex} & 0.0188 & 0.0147 & 0.0313 & 0.0179 & 0.0500 & 0.0232 & 0.0750 & 0.0299 & 0.0438 & 0.0227 & 0.0500 & 0.0243 \\
ENMF~\cite{chen2020enmf} & 0.0375 & 0.0261 & 0.0500 & 0.0292 & 0.0313 & 0.0185 & 0.0313 & 0.0185 & 0.0625 & 0.0319 & 0.0688 & 0.0336 \\
DiffRec~\cite{wang2023diffrec} & 0.0438 & 0.0199 & 0.0725 & 0.0267 & 0.0813 & 0.0480 & 0.1125 & 0.0587 & 0.0688 & 0.0331 & 0.1000 & 0.0408 \\
LRURec~\cite{yue2024linear} & 0.0463 & 0.0215 & 0.1013 & 0.0304 & 0.0650 & 0.0283 & 0.1025 & 0.0316 & 0.0550 & 0.0271 & 0.1138 & 0.0415 \\
FEARec~\cite{du2023fearec} & 0.0500 & 0.0189 & 0.1063 & 0.0331 & 0.0688 & 0.0305 & 0.1063 & 0.0339 & 0.0563 & 0.0283 & 0.1188 & 0.0439 \\ 
WGCL~\cite{chen2025squeeze} & 0.0525 & 0.0237 & 0.1088 & 0.0349 & 0.0850 & 0.0445 & 0.1175 & 0.0548 & 0.0650 & 0.0299 & 0.1038 & 0.0420 \\
NLGCL~\cite{xu2025nlgcl} & 0.0575 & 0.0259 & 0.1113 & 0.0367 & 0.0875 & 0.0475 & 0.1200 & 0.0574 & 0.0688 & 0.0317 & 0.1075 & 0.0435 \\
\midrule
BM25~\cite{robertson2009bm25} & 0.0063 & 0.0031 & 0.0063 & 0.0031 & 0.0000 & 0.0000 & 0.0000 & 0.0000 & 0.0000 & 0.0000 & 0.0063 & 0.0014 \\
LLMRank~\cite{hou2024llmrank} & 0.0125 & 0.0089 & 0.0125 & 0.0094 & 0.0000 & 0.0000 & 0.0063 & 0.0016 & 0.0125 & 0.0044 & 0.0125 & 0.0063 \\ 
\midrule
LightGCN$^{\mathrm{RAG}}$ & 0.0250 & 0.0141 & 0.0438 & 0.0189 & 0.0375 & 0.0185 & 0.0750 & 0.0347 & 0.0188 & 0.0120 & 0.0250 & 0.0126 \\
SASRec$^{\mathrm{RAG}}$ & 0.0625 & 0.0322 & 0.0750 & 0.0344 & 0.0625 & 0.0282 & 0.0750 & 0.0377 & 0.0188 & 0.0108 & 0.0188 & 0.0113 \\
SimpleX$^{\mathrm{RAG}}$ & 0.0438 & 0.0202 & 0.0563 & 0.0247 & 0.0500 & 0.0284 & 0.1000 & 0.0437 & 0.0063 & 0.0063 & 0.0188 & 0.0077 \\
ENMF$^{\mathrm{RAG}}$ & 0.0463 & 0.0287 & 0.0625 & 0.0331 & 0.0300 & 0.0172 & 0.0325 & 0.0188 & 0.0238 & 0.0119 & 0.0300 & 0.0125 \\
DiffRec$^{\mathrm{RAG}}$ & 0.0525 & 0.0231 & 0.0813 & 0.0297 & 0.0763 & 0.0458 & 0.1075 & 0.0561 & 0.0313 & 0.0148 & 0.0463 & 0.0189 \\
LRURec$^{\mathrm{RAG}}$ & 0.0513 & 0.0229 & 0.1125 & 0.0337 & 0.0613 & 0.0271 & 0.0963 & 0.0302 & 0.0263 & 0.0124 & 0.0363 & 0.0153 \\
FEARec$^{\mathrm{RAG}}$ & 0.0563 & 0.0215 & 0.1113 & 0.0340 & 0.0663 & 0.0291 & 0.1013 & 0.0331 & 0.0300 & 0.0136 & 0.0413 & 0.0185 \\
WGCL$^{\mathrm{RAG}}$ & 0.0613 & 0.0278 & 0.1113 & 0.0360 & 0.0813 & 0.0431 & 0.1125 & 0.0532 & 0.0313 & 0.0143 & 0.0463 & 0.0205 \\
NLGCL$^{\mathrm{RAG}}$ & 0.0663 & 0.0294 & 0.1125 & 0.0380 & 0.0838 & 0.0461 & 0.1163 & 0.0563 & 0.0338 & 0.0152 & 0.0488 & 0.0227 \\
\midrule
PRTA$^{\mathrm{Simple}}$~\cite{yang2026prta} & 0.0875 & 0.0416 & 0.1125 & 0.0542 & 0.1025 & 0.0554 & 0.1200 & 0.0583 & 0.1588 & 0.0751 & 0.1675 & 0.0809 \\
PRTA$^{\mathrm{Ultra}}$~\cite{yang2026prta} & \underline{0.1113} & \underline{0.0528} & \underline{0.1300} & \underline{0.0612} & \underline{0.1075} & 0.0583 & \underline{0.1263} & 0.0614 & \underline{0.1675} & 0.0791 & 0.1750 & 0.0834 \\
\midrule
BRAVE$^{\mathrm{Simple}}$ (ours) & 0.0950 & 0.0483 & 0.1175 & 0.0582 & 0.1063 & \underline{0.0602} & 0.1250 & \underline{0.0637} & 0.1663 & \underline{0.0820} & \underline{0.1825} & \underline{0.0854} \\
BRAVE$^{\mathrm{Ultra}}$ (ours) & \textbf{0.1138$^*$} & \textbf{0.0545$^*$} & \textbf{0.1338$^*$} & \textbf{0.0631$^*$} & \textbf{0.1113$^*$} & \textbf{0.0615$^*$} & \textbf{0.1300$^*$} & \textbf{0.0649$^*$} & \textbf{0.1713$^*$} & \textbf{0.0832$^*$} & \textbf{0.1875$^*$} & \textbf{0.0887$^*$} \\
\bottomrule
\end{tabular}}
% \vspace{-0.08in}
 \vskip -0.1in
\end{table*}

\subsection{Overall Performance}
\label{sec:overall}

\cref{tab:main-results} reports the performance of all compared methods on Amazon, Yelp, and Goodreads. We summarize three key observations.

\textbf{First, \brave achieves the best overall performance across datasets and metrics.}
Compared with conventional recommenders, language-only rankers, and RAG-style rerankers, \brave obtains stronger Recall@10, NDCG@10, Recall@20, and NDCG@20. The pattern indicates that recommendation agents benefit from learning personalized recommender-mixture preferences rather than relying on a single recommender, a text-only ranker, or a one-shot LLM reranking procedure. Among all methods, \braveUltra achieves the best results on all three datasets, with significant improvements over the strongest baseline where marked.

\textbf{Second, the improvement is not simply caused by a stronger recommender pool.}
Under the same recommender combination as PRTA, \braveSimple consistently outperforms PRTA$^{\mathrm{Simple}}$ across datasets and metrics. The gains are especially clear on NDCG: for example, Amazon NDCG@20 increases from 0.0542 to 0.0582, while Goodreads NDCG@20 increases from 0.0809 to 0.0854. The gain therefore comes from value-aware reflection allocation, not merely from replacing the underlying recommender tools. With the stronger recommender pool, \braveUltra further improves over PRTA$^{\mathrm{Ultra}}$ on all datasets, showing that budgeted allocation remains useful even when the base recommenders are already competitive.

\textbf{Third, \brave improves ranking quality in addition to hit coverage.}
The gains are especially clear on NDCG, which is sensitive to the ordering of relevant items near the top of the list. \braveUltra improves over PRTA$^{\mathrm{Ultra}}$ on NDCG@20 by 0.0019, 0.0035, and 0.0053 on Amazon, Yelp, and Goodreads, respectively. Selective reflection therefore appears to refine the learned recommender mixture and improve top-ranked ordering, rather than only expanding the hit set. Overall, full reflection is not necessary for every user: allocating reflection according to estimated user-level value can preserve or improve recommendation quality under a constrained reasoning budget.

\subsection{Ablation Study}
\label{sec:ablation}

\begin{figure}[t]
    \centering
    \includegraphics[width=.8\columnwidth]{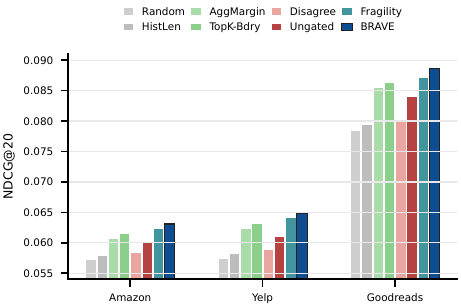}
     \vskip -0.1in
    \caption{Allocator ablation under the Ultra pool and 60\% reflection budget. Bars report NDCG@20.}
    \vskip -0.1in
    \label{fig:ablation}
\end{figure}

We compare \brave with cheap-selection baselines under the same budget. \textbf{Random} selects uniformly; \textbf{HistLen} prioritizes shorter histories; \textbf{Agg-margin} uses the top-1/top-2 margin of $S_u$; \textbf{Top-K boundary} uses the $K$-th/$(K+1)$-th aggregate margin; \textbf{Disagreement-only}, \textbf{Ungated additive}, and \textbf{Fragility-only} isolate the components of \cref{eq:gated}.

\cref{fig:ablation} shows that the gain is not explained by selecting arbitrary uncertain or cold-start users. Aggregate-boundary baselines are competitive, supporting the rank-flip analysis in \cref{sec:rankflip}, but still trail the gated allocator. Disagreement-only selection is unreliable, while fragility-only is stronger but below \brave. The key contrast is ungated additive versus \brave: both use the same signals, but ungated addition admits disagreement for every user, whereas \brave uses it only after the fragility gate.

\subsection{Reflection-Value Calibration}
\label{sec:value-calibration}

\begin{figure}[t]
    \centering
    \includegraphics[width=.8\columnwidth]{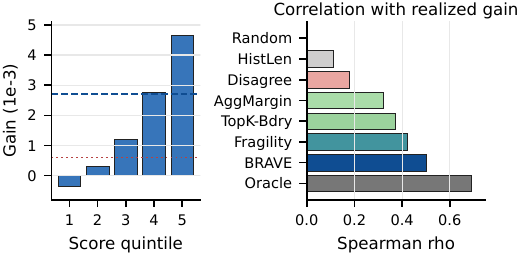}
     \vskip -0.1in
    \caption{Reflection-value calibration. Left: realized NDCG@20 gain by BRAVE-score quintile; dashed/dotted lines mark selected/unselected means. Right: Spearman correlation between each allocation score and realized gain.}
     \vskip -0.15in
    \label{fig:value-calibration}
\end{figure}

\cref{fig:value-calibration} tests whether the allocation score estimates the latent value in \cref{eq:value}. For each diagnostic-window user, we run matched branches from the same initial mixture memory and recommender outputs: one receives LLM local evaluation and global comparison, while the other receives only the rank-grounded update. Their NDCG@20 difference is the realized reflection gain. The left panel shows monotone gains across BRAVE-score quintiles: low-score users have near-zero or slightly negative gains, while the highest quintile contributes most improvement.

The right panel compares allocation scores against realized gain. History length and raw disagreement correlate weakly; aggregate-margin and top-$K$ boundary scores are stronger but remain below \brave. The oracle gap is expected because \brave uses only pre-reflection recommender evidence, yet selected users show substantially larger realized gain than unselected users.

\subsection{Cost-Quality Analysis}
\label{sec:cost-quality}
\begin{table}[t]
\centering
\caption{Amazon cost and diagnostic blocks. The budget block changes $B$, the allocator block changes only the selection rule, and the controller block changes only the LLM controller. All rows keep full-catalog ranking and inference reranking.}
 \vskip -0.1in
\label{tab:budget}
\scriptsize
\setlength{\tabcolsep}{1.8pt}
\resizebox{\columnwidth}{!}{
\begin{tabular}{llcccccc}
\toprule
Block & Setting & $B$ & Calls & R@10 & N@10 & R@20 & N@20 \\
\midrule
\multicolumn{8}{l}{\textit{Full reflection reference}} \\
Full & PRTA$^{\mathrm{Ultra}}$ & 100\% & 1.00$\times$ & 0.1113 & 0.0528 & 0.1300 & 0.0612 \\
\midrule
\multicolumn{8}{l}{\textit{Budget: vary the reflection budget with BRAVE allocation}} \\
Budget & BRAVE & 20\% & 0.31$\times$ & 0.1063 & 0.0506 & 0.1263 & 0.0588 \\
Budget & BRAVE & 40\% & 0.49$\times$ & 0.1125 & 0.0537 & 0.1325 & 0.0625 \\
Budget & BRAVE & 60\% & 0.66$\times$ & \textbf{0.1138} & \textbf{0.0545} & \textbf{0.1338} & \textbf{0.0631} \\
Budget & BRAVE & 80\% & 0.83$\times$ & 0.1138 & 0.0541 & 0.1325 & 0.0630 \\
\midrule
\multicolumn{8}{l}{\textit{Allocator: vary the selection rule at 60\% budget}} \\
Allocator & Random & 60\% & 0.66$\times$ & 0.1025 & 0.0489 & 0.1213 & 0.0572 \\
Allocator & Disagreement-only & 60\% & 0.66$\times$ & 0.1050 & 0.0498 & 0.1238 & 0.0584 \\
Allocator & Ungated additive & 60\% & 0.66$\times$ & 0.1088 & 0.0514 & 0.1275 & 0.0600 \\
Allocator & Fragility-only & 60\% & 0.66$\times$ & 0.1113 & 0.0532 & 0.1313 & 0.0623 \\
Allocator & BRAVE gated & 60\% & 0.66$\times$ & \textbf{0.1138} & \textbf{0.0545} & \textbf{0.1338} & \textbf{0.0631} \\
\midrule
\multicolumn{8}{l}{\textit{Controller: vary the LLM controller with BRAVE allocation}} \\
Closed ctrl. & GPT-4o & 60\% & 0.66$\times$ & 0.1150 & 0.0550 & 0.1350 & 0.0637 \\
Closed ctrl. & GPT-5.6 Luna & 60\% & 0.66$\times$ & 0.1138 & 0.0545 & 0.1338 & 0.0631 \\
Closed ctrl. & GPT-5.6 Terra & 60\% & 0.66$\times$ & \textbf{0.1163} & \textbf{0.0556} & \textbf{0.1363} & \textbf{0.0643} \\
Open ctrl. & Phi-4-mini & 60\% & 0.66$\times$ & 0.1125 & 0.0535 & 0.1325 & 0.0620 \\
Open ctrl. & Qwen3-8B & 60\% & 0.66$\times$ & 0.1125 & 0.0538 & 0.1325 & 0.0624 \\
Open ctrl. & GPT-OSS-20B & 60\% & 0.66$\times$ & 0.1138 & 0.0541 & 0.1325 & 0.0627 \\
\bottomrule
\end{tabular}}
 \vskip -0.1in
\end{table}

\begin{figure}[t]
    \centering
    \includegraphics[width=.75\columnwidth]{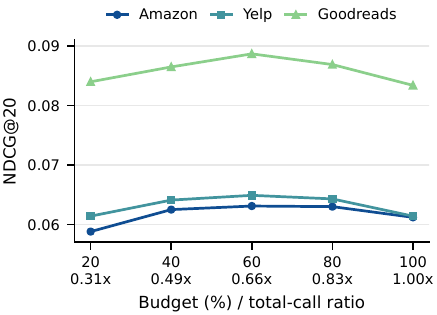}
     \vskip -0.1in
    \caption{Budget-quality trade-off. The second x-axis line reports the three-tool call ratio from \cref{eq:cost_ratio}.}
     \vskip -0.15in
    \label{fig:budget}
\end{figure}

\cref{tab:budget} contains three Amazon diagnostic blocks: the budget block varies how many users receive LLM reflection, the allocator block compares selection rules under the same 60\% budget, and the controller block changes only the LLM controller. Together with \cref{fig:budget}, the budget block shows a saturated cost-quality frontier. A 20\% budget loses useful updates, 40\% recovers most of the full-reflection benefit, and 60\% gives the best top-ranked quality. Moving to 80\% offers little gain, showing that more reflection is not automatically better when extra calls go to stable or unhelpfully conflicting users. The diagnostic rows in \cref{tab:budget} support the same pattern: random and disagreement-only selection are weaker, fragility-only is competitive, and stronger LLMs improve absolute scores without removing the need for allocation. \brave thus keeps the reflection operation unchanged but moves it toward users with higher expected value.

\subsection{Boundary-Swap and Disagreement Calibration}
\label{sec:allocator-analysis}

\begin{figure}[t]
    \centering
    \includegraphics[width=.8\columnwidth]{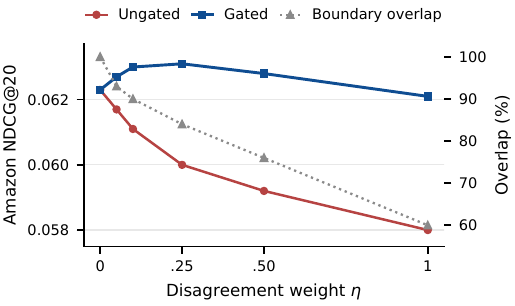}
     \vskip -0.1in
    \caption{Amazon-only disagreement calibration. Gating keeps cross-tool disagreement conditional on ranking fragility; the gray curve shows selected-set overlap with the fragility-only boundary set.}
     \vskip -0.15in
    \label{fig:disagreement}
\end{figure}

We test the distinction between ranking fragility and cross-tool disagreement. \cref{fig:disagreement} varies $\eta$ and compares ungated addition with gated estimator in \cref{eq:gated}. Ungated NDCG@20 decreases as $\eta$ grows: high disagreement can pull the allocator toward users whose recommender tools disagree even when reflection is unlikely to improve the final ranking. \brave is more stable because disagreement affects priority only after the fragility gate. The selected-set overlap curve explains the mechanism. As $\eta$ increases, disagreement-aware selection diverges from the fragility boundary set; replacing fragile high-value users with high-disagreement low-value users can reduce ranking quality. This boundary-swap effect supports using disagreement to refine, not define, reflection opportunity.

\subsection{Hyperparameter Sensitivity}
\label{sec:hyperparameter}

\begin{figure}[t]
    \centering
    \includegraphics[width=.8\columnwidth]{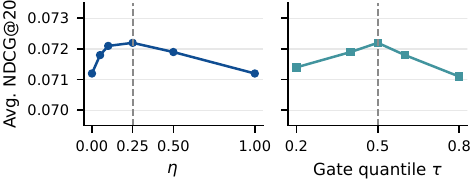}
     \vskip -0.1in
    \caption{Sensitivity to $\eta$ and $\tau$. Values are NDCG@20 averaged over the three datasets under the Ultra pool; dashed lines mark defaults.}
     \vskip -0.1in
    \label{fig:hyperparameter}
\end{figure}

\cref{fig:hyperparameter} shows stable performance near the default. Unlike the Amazon-only diagnostic in \cref{fig:disagreement}, this figure averages over Amazon, Yelp, and Goodreads. Very small $\eta$ approaches fragility-only selection, while very large $\eta$ overweights disagreement and weakens ranking quality. The gate quantile $\tau$ behaves similarly: loose gates admit disagreement too broadly, whereas strict gates suppress useful disagreement. The default $\eta=0.25$ and $\tau=0.5$ are robust operating points rather than narrow optima.

\subsection{Tool-Pool Composition}
\label{sec:tool-pool}

\begin{figure}[t]
    \centering
    \includegraphics[width=.9\columnwidth]{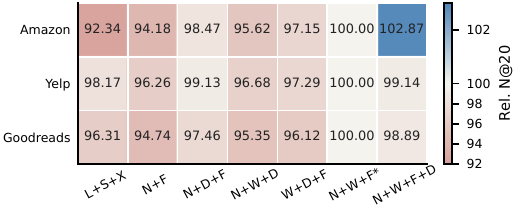}
     \vskip -0.1in
    \caption{Tool-pool composition study. Each cell reports NDCG@20 normalized by the selected Ultra pool on the same dataset. L/S/X/N/W/F/D denote LightGCN/SASRec/SimpleX/NLGCL/WGCL/FEARec/DiffRec; * marks the Ultra pool.}
     \vskip -0.15in
    \label{fig:tool-pool}
\end{figure}

\cref{fig:tool-pool} studies how the Ultra pool is selected. We compare the PRTA Simple pool, the strongest compact two-recommender pool (NLGCL + FEARec), three substitution pools, the selected Ultra pool (NLGCL + WGCL + FEARec), and a four-recommender expansion with DiffRec. All variants keep the \brave allocator, budget, controller, and inference reranking fixed. Three patterns emerge. Simple is insufficient for the stronger setting, so budgeted reflection still depends on high-quality recommender evidence. NLGCL + FEARec is strong but drops about 3--6\% from Ultra, indicating that WGCL adds complementary graph-neighborhood evidence. Substitution is not interchangeable: replacing WGCL with DiffRec, removing FEARec, or removing NLGCL weakens at least one dataset. A fourth recommender tool helps Amazon by about 3\% but reduces Yelp and Goodreads by about 1\%, suggesting that extra evidence can also introduce noisy disagreement. We therefore use NLGCL + WGCL + FEARec as the default Ultra pool.

\subsection{Controller Robustness}
\label{sec:controller-robustness}

\begin{figure}[t]
    \centering
    \includegraphics[width=0.8\columnwidth]{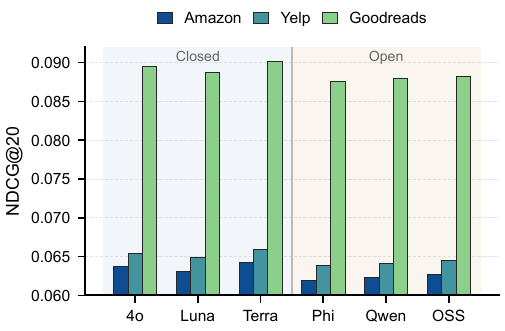}
     \vskip -0.1in
    \caption{Controller Analysis. Only LLM controller changes.}
     \vskip -0.15in
    \label{fig:controller}
\end{figure}

\cref{fig:controller} tests whether the policy is tied to one controller. We compare closed-source controllers (GPT-4o, GPT-5.6 Luna, GPT-5.6 Terra) with open-source controllers (Phi-4-mini, Qwen3-8B, GPT-OSS-20B), keeping prompts, recommender outputs, and allocation fixed. Larger closed-source controllers give modest gains over GPT-5.6 Luna, while open-source controllers incur only slight drops. Thus the allocator benefits from stronger controllers but does not require the strongest available model.

% \FloatBarrier
\section{Conclusion}
This paper studies reflection allocation for recommendation agents. Instead of assuming that every user should receive the same amount of LLM reflection, we argue that reflection has user-dependent value and should be spent where it is most likely to change the learned recommender mixture. We propose \brave, a budgeted reflection-allocation framework that estimates reflection value from recommender-side self-doubt before making training-time LLM calls. By gating cross-tool disagreement with ranking fragility, \brave preserves the base agent's recommender tools, prompts, and inference reranking while selectively reducing repeated reflection.

Our analysis explains why low-margin users provide greater rank-flip opportunity and why disagreement should be treated as a conditional signal rather than an unconditional objective. Experiments on full-ranking recommendation benchmarks show that \brave improves the cost-quality trade-off over uniform reflection and remains effective across tool pools and LLM controllers. These results suggest that scalable recommendation agents require not only stronger reflection modules, but also principled policies for deciding when reflection is worth using.

% \section*{GenAI Usage Disclosure}
% Generative AI tools were used to assist with language drafting and editing. The research idea, method design, experimental protocol, result analysis, and final manuscript remain the responsibility of the authors.

\subsection*{Ethical Considerations}
This work aims to make recommendation agents more efficient by reducing unnecessary LLM reflection while preserving full-catalog ranking and inference reranking. Since \brave changes only how training-time reflection is allocated, it does not introduce new user data requirements or additional LLM exposure beyond the base agent protocol. The main potential risk is that a poorly calibrated allocator may assign less reflection to users whose preferences are sparse, niche, or under-represented, which could weaken personalization for those users or indirectly affect item exposure.

This risk can be mitigated through deployment-time auditing and conservative allocation design. In practice, reflection budgets should be evaluated not only by aggregate Recall or NDCG, but also across user-history lengths, item popularity groups, and application-relevant cohorts when such analysis is permitted. The allocator should rely on task-side ranking evidence rather than sensitive attributes, and fallback policies can reserve a small portion of the budget for random or fairness-aware coverage when subgroup robustness is important. Privacy risks can be controlled with the same protections required by the base recommendation agent: minimizing user text sent to the LLM controller, anonymizing user identifiers, retaining only necessary interaction evidence, and applying standard access-control and logging policies. Overall, \brave is designed as a cost-aware allocation layer, and its ethical deployment depends on validating that the saved computation does not come at the expense of vulnerable users or responsible exposure.

\balance
\bibliographystyle{ACM-Reference-Format}
\bibliography{reference}

@inproceedings{bao2023tallrec,
  title = {TALLRec: An Effective and Efficient Tuning Framework to Align Large Language Model with Recommendation},
  author = {Bao, Keqin and Zhang, Jizhi and Zhang, Yang and Wang, Wenjie and Feng, Fuli and He, Xiangnan},
  booktitle = {Proceedings of the 17th ACM Conference on Recommender Systems},
  pages = {1007--1014},
  year = {2023},
  doi = {10.1145/3604915.3608857}
}

@article{chen2020enmf,
  title = {Efficient Neural Matrix Factorization without Sampling for Recommendation},
  author = {Chen, Chong and Zhang, Min and Zhang, Yongfeng and Liu, Yiqun and Ma, Shaoping},
  journal = {ACM Transactions on Information Systems},
  volume = {38},
  number = {2},
  pages = {14:1--14:28},
  year = {2020},
  doi = {10.1145/3373807}
}

@article{chen2023frugalgpt,
  title = {FrugalGPT: How to Use Large Language Models While Reducing Cost and Improving Performance},
  author = {Chen, Lingjiao and Zaharia, Matei and Zou, James},
  journal = {CoRR},
  volume = {abs/2305.05176},
  year = {2023},
  doi = {10.48550/arXiv.2305.05176}
}

@inproceedings{du2023fearec,
  title = {Frequency Enhanced Hybrid Attention Network for Sequential Recommendation},
  author = {Du, Xinyu and Yuan, Huanhuan and Zhao, Pengpeng and Qu, Jianfeng and Zhuang, Fuzhen and Liu, Guanfeng and Liu, Yanchi and Sheng, Victor S.},
  booktitle = {Proceedings of the 46th International ACM SIGIR Conference on Research and Development in Information Retrieval},
  pages = {78--88},
  year = {2023},
  doi = {10.1145/3539618.3591689}
}

@article{xu2026beyond,
  title={Beyond Successor Accuracy: State Retention for Recursive Self-Improvement in Recommendation},
  author={Xu, Jinfeng and Chen, Zheyu and Peng, Ziyue and Lin, Zheng and Yuan, Wenhao and Chen, Jian and Li, Shujie and Ngai, Edith},
  journal={arXiv preprint arXiv:2610.07105},
  year={2026}
}

@inproceedings{xu2026iimrec,
  title={One Graph, Multiple Gains: Single High-Quality Item-Item Graph for Multimodal Recommendation},
  author={Xu, Jinfeng and Chen, Zheyu and Peng, Ziyue and Yang, Shuo and Li, Jinze and Liu, Zewei and Li, Shujie and Du, Yipeng and Ngai, Edith C. H.},
  booktitle={Proceedings of the 34th ACM International Conference on Multimedia},
  pages={12508--12516},
  year={2026}
}

@inproceedings{xu2026ntcf,
  title={Neural Tree Collaborative Filtering: Rethinking Graph Collaborative Filtering as Tree Collaborative Filtering with Curvature-Aware Propagation Depth},
  author={Xu, Jinfeng and Chen, Zheyu and Peng, Ziyue and Yang, Shuo and Li, Jinze and Yuan, Wenhao and Chen, Jian and Li, Shujie and Ngai, Edith C. H.},
  booktitle={Proceedings of the 35th ACM International Conference on Information and Knowledge Management},
  pages={7991--7996},
  year={2026}
}

@inproceedings{geng2022p5,
  title = {Recommendation as Language Processing (RLP): A Unified Pretrain, Personalized Prompt and Predict Paradigm (P5)},
  author = {Geng, Shijie and Liu, Shuchang and Fu, Zuohui and Ge, Yingqiang and Zhang, Yongfeng},
  booktitle = {Proceedings of the 16th ACM Conference on Recommender Systems},
  pages = {299--315},
  year = {2022},
  doi = {10.1145/3523227.3546767}
}

@inproceedings{he2020lightgcn,
  title = {LightGCN: Simplifying and Powering Graph Convolution Network for Recommendation},
  author = {He, Xiangnan and Deng, Kuan and Wang, Xiang and Li, Yan and Zhang, Yongdong and Wang, Meng},
  booktitle = {Proceedings of the 43rd International ACM SIGIR Conference on Research and Development in Information Retrieval},
  pages = {639--648},
  year = {2020},
  doi = {10.1145/3397271.3401063}
}

@inproceedings{geifman2017selective,
  title = {Selective Classification for Deep Neural Networks},
  author = {Geifman, Yonatan and El-Yaniv, Ran},
  booktitle = {Advances in Neural Information Processing Systems},
  volume = {30},
  year = {2017}
}

@inproceedings{geifman2019selectivenet,
  title = {SelectiveNet: A Deep Neural Network with an Integrated Reject Option},
  author = {Geifman, Yonatan and El-Yaniv, Ran},
  booktitle = {Proceedings of the 36th International Conference on Machine Learning},
  pages = {2151--2159},
  year = {2019},
  url = {https://proceedings.mlr.press/v97/geifman19a.html}
}

@article{graves2016act,
  title = {Adaptive Computation Time for Recurrent Neural Networks},
  author = {Graves, Alex},
  journal = {CoRR},
  volume = {abs/1603.08983},
  year = {2016},
  doi = {10.48550/arXiv.1603.08983}
}

@inproceedings{hou2024llmrank,
  title = {Large Language Models are Zero-Shot Rankers for Recommender Systems},
  author = {Hou, Yupeng and Zhang, Junjie and Lin, Zihan and Lu, Hongyu and Xie, Ruobing and McAuley, Julian J. and Zhao, Wayne Xin},
  booktitle = {Advances in Information Retrieval},
  pages = {364--381},
  year = {2024},
  doi = {10.1007/978-3-031-56060-6_24}
}

@article{ji2023hallucination,
  title = {Survey of Hallucination in Natural Language Generation},
  author = {Ji, Ziwei and Lee, Nayeon and Frieske, Rita and Yu, Tiezheng and Su, Dan and Xu, Yan and Ishii, Etsuko and Bang, Yejin and Madotto, Andrea and Fung, Pascale},
  journal = {ACM Computing Surveys},
  volume = {55},
  number = {12},
  pages = {248:1--248:38},
  year = {2023},
  doi = {10.1145/3571730}
}

@inproceedings{jiang2025beyond,
  title = {Beyond Utility: Evaluating LLM as Recommender},
  author = {Jiang, Chumeng and Wang, Jiayin and Ma, Weizhi and Clarke, Charles L. A. and Wang, Shuai and Wu, Chuhan and Zhang, Min},
  booktitle = {Proceedings of the ACM Web Conference 2025},
  pages = {3850--3862},
  year = {2025},
  doi = {10.1145/3696410.3714759}
}

@inproceedings{kang2018sasrec,
  title = {Self-Attentive Sequential Recommendation},
  author = {Kang, Wang-Cheng and McAuley, Julian J.},
  booktitle = {IEEE International Conference on Data Mining},
  pages = {197--206},
  year = {2018},
  doi = {10.1109/ICDM.2018.00035}
}

@article{karpas2022mrkl,
  title = {MRKL Systems: A Modular, Neuro-Symbolic Architecture that Combines Large Language Models, External Knowledge Sources and Discrete Reasoning},
  author = {Karpas, Ehud and Abend, Omri and Belinkov, Yonatan and Lenz, Barak and Lieber, Opher and Ratner, Nir and Shoham, Yoav and Bata, Hofit and Levine, Yoav and Leyton-Brown, Kevin and Muhlgay, Dor and Rozen, Noam and Schwartz, Erez and Shachaf, Gal and Shalev-Shwartz, Shai and Shashua, Amnon and Tenenholtz, Moshe},
  journal = {CoRR},
  volume = {abs/2205.00445},
  year = {2022},
  doi = {10.48550/arXiv.2205.00445}
}

@inproceedings{lewis1994uncertainty,
  title = {A Sequential Algorithm for Training Text Classifiers},
  author = {Lewis, David D. and Gale, William A.},
  booktitle = {Proceedings of the 17th Annual International ACM SIGIR Conference on Research and Development in Information Retrieval},
  pages = {3--12},
  year = {1994},
  doi = {10.1007/978-1-4471-2099-5_1}
}

@inproceedings{mao2021simplex,
  title = {SimpleX: A Simple and Strong Baseline for Collaborative Filtering},
  author = {Mao, Kelong and Zhu, Jieming and Wang, Jinpeng and Dai, Quanyu and Dong, Zhenhua and Xiao, Xi and He, Xiuqiang},
  booktitle = {Proceedings of the 30th ACM International Conference on Information and Knowledge Management},
  pages = {1243--1252},
  year = {2021},
  doi = {10.1145/3459637.3482297}
}

@inproceedings{madaan2023selfrefine,
  title = {Self-Refine: Iterative Refinement with Self-Feedback},
  author = {Madaan, Aman and Tandon, Niket and Gupta, Prakhar and Hallinan, Skyler and Gao, Luyu and Wiegreffe, Sarah and Alon, Uri and Dziri, Nouha and Prabhumoye, Shrimai and Yang, Yiming and Gupta, Shashank and Majumder, Bodhisattwa Prasad and Hermann, Katherine and Welleck, Sean and Yazdanbakhsh, Amir and Clark, Peter},
  booktitle = {Advances in Neural Information Processing Systems},
  volume = {36},
  year = {2023}
}

@inproceedings{ni2019justifying,
  title = {Justifying Recommendations using Distantly-Labeled Reviews and Fine-Grained Aspects},
  author = {Ni, Jianmo and Li, Jiacheng and McAuley, Julian J.},
  booktitle = {Proceedings of the 2019 Conference on Empirical Methods in Natural Language Processing and the 9th International Joint Conference on Natural Language Processing},
  pages = {188--197},
  year = {2019},
  doi = {10.18653/v1/D19-1018}
}

@inproceedings{patil2024gorilla,
  title = {Gorilla: Large Language Model Connected with Massive APIs},
  author = {Patil, Shishir G. and Zhang, Tianjun and Wang, Xin and Gonzalez, Joseph E.},
  booktitle = {Advances in Neural Information Processing Systems},
  volume = {37},
  year = {2024}
}

@article{robertson2009bm25,
  title = {The Probabilistic Relevance Framework: BM25 and Beyond},
  author = {Robertson, Stephen E. and Zaragoza, Hugo},
  journal = {Foundations and Trends in Information Retrieval},
  volume = {3},
  number = {4},
  pages = {333--389},
  year = {2009},
  doi = {10.1561/1500000019}
}

@inproceedings{schick2023toolformer,
  title = {Toolformer: Language Models Can Teach Themselves to Use Tools},
  author = {Schick, Timo and Dwivedi-Yu, Jane and Dessi, Roberto and Raileanu, Roberta and Lomeli, Maria and Hambro, Eric and Zettlemoyer, Luke and Cancedda, Nicola and Scialom, Thomas},
  booktitle = {Advances in Neural Information Processing Systems},
  year = {2023}
}

@inproceedings{seung1992query,
  title = {Query by Committee},
  author = {Seung, H. Sebastian and Opper, Manfred and Sompolinsky, Haim},
  booktitle = {Proceedings of the Fifth Annual Workshop on Computational Learning Theory},
  pages = {287--294},
  year = {1992},
  doi = {10.1145/130385.130417}
}

@techreport{settles2009activelearning,
  title = {Active Learning Literature Survey},
  author = {Settles, Burr},
  institution = {University of Wisconsin-Madison},
  number = {1648},
  year = {2009}
}

@inproceedings{shinn2023reflexion,
  title = {Reflexion: Language Agents with Verbal Reinforcement Learning},
  author = {Shinn, Noah and Cassano, Federico and Gopinath, Ashwin and Narasimhan, Karthik and Yao, Shunyu},
  booktitle = {Advances in Neural Information Processing Systems},
  volume = {36},
  year = {2023}
}

@inproceedings{shen2023hugginggpt,
  title = {HuggingGPT: Solving AI Tasks with ChatGPT and its Friends in Hugging Face},
  author = {Shen, Yongliang and Song, Kaitao and Tan, Xu and Li, Dongsheng and Lu, Weiming and Zhuang, Yueting},
  booktitle = {Advances in Neural Information Processing Systems},
  volume = {36},
  year = {2023}
}

@article{tong2001svm,
  title = {Support Vector Machine Active Learning with Applications to Text Classification},
  author = {Tong, Simon and Koller, Daphne},
  journal = {Journal of Machine Learning Research},
  volume = {2},
  pages = {45--66},
  year = {2001}
}

@inproceedings{wang2024llmnotfair,
  title = {Large Language Models are not Fair Evaluators},
  author = {Wang, Peiyi and Li, Lei and Chen, Liang and Cai, Zefan and Zhu, Dawei and Lin, Binghuai and Cao, Yunbo and Kong, Lingpeng and Liu, Qi and Liu, Tianyu and Sui, Zhifang},
  booktitle = {Proceedings of the 62nd Annual Meeting of the Association for Computational Linguistics (Volume 1: Long Papers)},
  pages = {9440--9450},
  year = {2024},
  doi = {10.18653/v1/2024.acl-long.511}
}

@article{wang2025simulation,
  title = {User Behavior Simulation with Large Language Model-based Agents},
  author = {Wang, Lei and Zhang, Jingsen and Yang, Hao and Chen, Zhi-Yuan and Tang, Jiakai and Zhang, Zeyu and Chen, Xu and Lin, Yankai and Sun, Hao and Song, Ruihua and Zhao, Xin and Xu, Jun and Dou, Zhicheng and Wang, Jun and Wen, Ji-Rong},
  journal = {ACM Transactions on Information Systems},
  volume = {43},
  number = {2},
  pages = {55:1--55:37},
  year = {2025},
  doi = {10.1145/3708985}
}

@inproceedings{wan2018item,
  title = {Item Recommendation on Monotonic Behavior Chains},
  author = {Wan, Mengting and McAuley, Julian},
  booktitle = {Proceedings of the 12th ACM Conference on Recommender Systems},
  pages = {86--94},
  year = {2018},
  doi = {10.1145/3240323.3240369}
}

@inproceedings{wang2023diffrec,
  title = {Diffusion Recommender Model},
  author = {Wang, Wenjie and Xu, Yiyan and Feng, Fuli and Lin, Xinyu and He, Xiangnan and Chua, Tat-Seng},
  booktitle = {Proceedings of the 46th International ACM SIGIR Conference on Research and Development in Information Retrieval},
  pages = {832--841},
  year = {2023},
  doi = {10.1145/3539618.3591663}
}

@inproceedings{wu2024avatar,
  title = {AvaTaR: Optimizing LLM Agents for Tool Usage via Contrastive Reasoning},
  author = {Wu, Shirley and Zhao, Shiyu and Huang, Qian and Huang, Kexin and Yasunaga, Michihiro and Cao, Kaidi and Ioannidis, Vassilis N. and Subbian, Karthik and Leskovec, Jure and Zou, James Y.},
  booktitle = {Advances in Neural Information Processing Systems},
  year = {2024}
}

@inproceedings{xu2024selfbias,
  title = {Pride and Prejudice: LLM Amplifies Self-Bias in Self-Refinement},
  author = {Xu, Wenda and Zhu, Guanglei and Zhao, Xuandong and Pan, Liangming and Li, Lei and Wang, William Yang},
  booktitle = {Proceedings of the 62nd Annual Meeting of the Association for Computational Linguistics (Volume 1: Long Papers)},
  pages = {15474--15492},
  year = {2024},
  doi = {10.18653/v1/2024.acl-long.826}
}

@inproceedings{xin2020deebert,
  title = {DeeBERT: Dynamic Early Exiting for Accelerating BERT Inference},
  author = {Xin, Ji and Tang, Raphael and Lee, Jaejun and Yu, Yaoliang and Lin, Jimmy},
  booktitle = {Proceedings of the 58th Annual Meeting of the Association for Computational Linguistics},
  pages = {2246--2251},
  year = {2020},
  doi = {10.18653/v1/2020.acl-main.204}
}

@misc{yelpdataset,
  title = {Yelp Open Dataset},
  author = {{Yelp}},
  year = {2024},
  url = {https://www.yelp.com/dataset}
}

@inproceedings{yang2026agentdr,
  title = {AgentDR: Dynamic Recommendation with Implicit Item-Item Relations via LLM-based Agents},
  author = {Yang, Mingdai and Choudhary, Nurendra and Du, Jiangshu and Huang, Edward W. and Yu, Philip and Subbian, Karthik and Koutra, Danai},
  booktitle = {Proceedings of the ACM Web Conference 2026},
  pages = {6230--6241},
  year = {2026},
  doi = {10.1145/3774904.3792304}
}

@inproceedings{yu2026thinkrec,
  title = {ThinkRec: Thinking-based Recommendation via LLM},
  author = {Yu, Qihang and Fu, Kairui and Lv, Zheqi and Zhang, Shengyu and Wu, Xinhui and Lin, Chen and Wei, Feng and Zheng, Bo and Wu, Fei},
  booktitle = {Proceedings of the ACM Web Conference 2026},
  pages = {5698--5709},
  year = {2026},
  doi = {10.1145/3774904.3792070}
}

@inproceedings{yang2026prta,
  title = {Personalized Recommendation Tool Learning via Autonomous Language Agents},
  author = {Yang, Mingdai and Liu, Zhiwei and Zhang, Weizhi and Wang, Yibo and Peng, Hao and Yu, Philip S.},
  booktitle = {Proceedings of the 20th ACM Conference on Recommender Systems},
  pages = {1--6},
  year = {2026},
  doi = {10.1145/3773078.3831794}
}

@inproceedings{yao2023react,
  title = {ReAct: Synergizing Reasoning and Acting in Language Models},
  author = {Yao, Shunyu and Zhao, Jeffrey and Yu, Dian and Du, Nan and Shafran, Izhak and Narasimhan, Karthik R. and Cao, Yuan},
  booktitle = {International Conference on Learning Representations},
  year = {2023},
  url = {https://openreview.net/forum?id=WE_vluYUL-X}
}

@inproceedings{yue2024llmcascade,
  title = {Large Language Model Cascades with Mixture of Thought Representations for Cost-Efficient Reasoning},
  author = {Yue, Murong and Zhao, Jie and Zhang, Min and Du, Liang and Yao, Ziyu},
  booktitle = {International Conference on Learning Representations},
  year = {2024}
}

@inproceedings{zhang2024agentcf,
  title = {AgentCF: Collaborative Learning with Autonomous Language Agents for Recommender Systems},
  author = {Zhang, Junjie and Hou, Yupeng and Xie, Ruobing and Sun, Wenqi and McAuley, Julian J. and Zhao, Wayne Xin and Lin, Leyu and Wen, Ji-Rong},
  booktitle = {Proceedings of the ACM Web Conference 2024},
  pages = {3679--3689},
  year = {2024},
  doi = {10.1145/3589334.3645537}
}

@inproceedings{zhang2024generativeagents,
  title = {On Generative Agents in Recommendation},
  author = {Zhang, An and Chen, Yuxin and Sheng, Leheng and Wang, Xiang and Chua, Tat-Seng},
  booktitle = {Proceedings of the 47th International ACM SIGIR Conference on Research and Development in Information Retrieval},
  pages = {1807--1817},
  year = {2024},
  doi = {10.1145/3626772.3657844}
}

@inproceedings{zhou2024lats,
  title = {Language Agent Tree Search Unifies Reasoning, Acting, and Planning in Language Models},
  author = {Zhou, Andy and Yan, Kai and Shlapentokh-Rothman, Michal and Wang, Haohan and Wang, Yu-Xiong},
  booktitle = {Proceedings of the 41st International Conference on Machine Learning},
  pages = {62138--62160},
  year = {2024}
}

@inproceedings{zhao2024toolrec,
  title = {Let Me Do It For You: Towards LLM Empowered Recommendation via Tool Learning},
  author = {Zhao, Yuyue and Wu, Jiancan and Wang, Xiang and Tang, Wei and Wang, Dingxian and de Rijke, Maarten},
  booktitle = {Proceedings of the 47th International ACM SIGIR Conference on Research and Development in Information Retrieval},
  pages = {1796--1806},
  year = {2024},
  doi = {10.1145/3626772.3657828}
}

@inproceedings{xu2025nlgcl,
  title={NLGCL: Naturally Existing Neighbor Layers Graph Contrastive Learning for Recommendation},
  author={Xu, Jinfeng and Chen, Zheyu and Yang, Shuo and Li, Jinze and Wang, Hewei and Wang, Wei and Hu, Xiping and Ngai, Edith},
  booktitle={Proceedings of the Nineteenth ACM Conference on Recommender Systems},
  pages={319--329},
  year={2025}
}

@inproceedings{chen2025squeeze,
  title={Squeeze and Excitation: A Weighted Graph Contrastive Learning for Collaborative Filtering},
  author={Chen, Zheyu and Xu, Jinfeng and Wei, Yutong and Peng, Ziyue},
  booktitle={Proceedings of the 48th International ACM SIGIR Conference on Research and Development in Information Retrieval},
  pages={2769--2773},
  year={2025}
}

@inproceedings{yue2024linear,
  title={Linear recurrent units for sequential recommendation},
  author={Yue, Zhenrui and Wang, Yueqi and He, Zhankui and Zeng, Huimin and McAuley, Julian and Wang, Dong},
  booktitle={Proceedings of the 17th ACM international conference on web search and data mining},
  pages={930--938},
  year={2024}
}

@misc{openai2026gpt56,
  author       = {{OpenAI}},
  title        = {{GPT-5.6: Frontier Intelligence That Scales with Your Ambition}},
  year         = {2026},
  month        = jul,
  howpublished = {\url{https://openai.com/index/gpt-5-6/}},
  note         = {Accessed: 2026-08-02}
}

@inproceedings{chen2025hypercomplex,
  title={Hypercomplex Prompt-aware Multimodal Recommendation},
  author={Chen, Zheyu and Xu, Jinfeng and Wang, Hewei and Yang, Shuo and Wan, Zitong and Hu, Haibo},
  booktitle={Proceedings of the 34th ACM International Conference on Information and Knowledge Management},
  pages={403--414},
  year={2025}
}

@article{xu2025survey,
  title={A survey on multimodal recommender systems: Recent advances and future directions},
  author={Xu, Jinfeng and Chen, Zheyu and Yang, Shuo and Li, Jinze and Wang, Wei and Hu, Xiping and Hoi, Steven and Ngai, Edith},
  journal={IEEE Transactions on Multimedia},
  year={2026},
  publisher={IEEE}
}

\end{document}